\documentclass[a4paper,11pt]{article}
\usepackage{pos}

\usepackage{colortbl}

\title{Feynman integrals, geometries and differential equations}

\author[a]{Sebastian P\"ogel}
\author[b]{Xing Wang}
\author*[a]{Stefan Weinzierl}

\affiliation[a]{
 PRISMA Cluster of Excellence, 
 Institut f{\"u}r Physik, Staudinger Weg 7, \\
 Johannes Gutenberg-Universit{\"a}t Mainz, 
 D - 55099 Mainz, Germany
}

\affiliation[b]{
 Technische Universit\"at M\"unchen, 
 TUM School of Natural Sciences, 
 Physik Department,\\
 D- 85748 Garching, Germany
}

\emailAdd{poegel@uni-mainz.de}
\emailAdd{xing.wang@tum.de}
\emailAdd{weinzierl@uni-mainz.de}

\abstract{
In this talk we discuss the construction of a basis of master integrals for the family of the
$l$-loop equal-mass banana integrals, such that the differential equation is in an $\varepsilon$-factorised form.
As the $l$-loop banana integral is related to a Calabi-Yau $(l-1)$-fold, 
this extends the examples where an $\varepsilon$-factorised form
has been found from Feynman integrals related to curves (of genus zero and one) 
to Feynman integrals related to higher-dimensional varieties.
}

\FullConference{16th International Symposium on Radiative Corrections: Applications of Quantum Field Theory to Phenomenology (RADCOR2023)\\
28th May - 2nd June, 2023\\
Crieff, Scotland, UK\\}

\newcommand{\bq}{\begin{eqnarray}}
\newcommand{\eq}{\end{eqnarray}}
\newcommand{\eps}{\varepsilon}

\newcommand{\qbar}{q}

\newcommand{\NB}{N_B}
\newcommand{\NF}{N_F}
\newcommand{\NL}{N_L}

\newcommand{\Frobeniusbasis}{\psi}
\newcommand{\Yinvariant}{Y}

\begin{document}
\maketitle


\section{Introduction}

Feynman integrals are indispensable for precision calculations in quantum field theory.
We would like to make precise predictions for observables in scattering experiments.
Any such calculation will involve a scattering amplitude.
Unfortunately we cannot calculate scattering amplitudes exactly.
If we have a small parameter like a small coupling, we may use perturbation theory. 
We may organise the perturbative expansion of a scattering amplitude in terms of Feynman diagrams, such that the
scattering amplitude is given by the sum of all relevant Feynman diagrams.
It is then a natural question to ask, what special functions appear in the final answer for the scattering amplitude.
Although there are situations, where the final answer for a scattering amplitude can be written in a rather compact form
due to significant simplifications after the sum over all relevant Feynman diagrams is taken, the class of functions in which the result
is expressed usually stays the same.
It is therefore sensible to ask in which class of functions an individual Feynman diagram can be expressed.
The answer to this question reveals a deep connection between Feynman integrals, geometry and differential equations.

The simplest Feynman integrals can be expressed in terms of multiple polylogarithms.
From the mathematical side, multiple polylogarithms are iterated integrals on a curve of genus zero with a certain number of marked points.
The next more complicated Feynman integrals are related to a curve of genus one with a certain number of marked points.
These are known as elliptic Feynman integrals, and have received significant attention in the last ten years.
There are two paths for further generalisations:
On the one hand we may go to curves of higher genus \cite{Huang:2013kh,Georgoudis:2015hca,Doran:2023yzu,Marzucca:2023gto},
on the other hand we may go to varieties of higher dimension.
The latter case is already required for rather simple Feynman integrals, as for example the family of banana graphs.
In this talk we put an emphasis on Calabi-Yau geometries \cite{Bloch:2014qca,Bloch:2016izu,Bourjaily:2018ycu,Bourjaily:2018yfy,Bourjaily:2019hmc,Klemm:2019dbm,Vergu:2020uur,Bonisch:2020qmm,Bonisch:2021yfw,Pogel:2022yat,Pogel:2022ken,Pogel:2022vat,Duhr:2022pch,Duhr:2022dxb,Kreimer:2022fxm,Forum:2022lpz,Cao:2023tpx,McLeod:2023doa}, which are generalisations of elliptic curves (one-dimensional varieties)
to higher dimensions.

A standard technique for the computation of Feynman integrals is the method of differential equations.
As it is common practice in our field, we use dimensional regularisation 
to regulate ultraviolet and infrared divergences. We set the number of space-time dimensions to $D=D_{\mathrm{int}}-2\eps$,
where $D_{\mathrm{int}}$ is the integer number of space-time dimensions we are interested in and
$\eps$ is the dimensional regularisation parameter.
Integration-by-parts identities \cite{Chetyrkin:1981qh} allow us to express any Feynman integral
from a family of Feynman integrals as a finite linear combination of a subset of this family.
The integrals of this subset are called master integrals and define a basis of a vector space. 
We denote the master integrals by $I=(I_1,I_2,\dots,I_{\NF})$.
In particular, we may express the derivatives of the master integrals with respect to the kinematic variables again as a linear
combination of the master integrals.
This leads to the 
differential equation \cite{Kotikov:1990kg}
\bq
 d I & = & A\left(x,\eps\right) I.
\eq
Let us stress that there are no conceptional issues in deriving the differential equation.
It only involves linear algebra and is always possible.
However, there can be practical problems, if the size of the linear system gets too large.
This reduces the problem of computing a Feynman integral to the problem of solving a system of differential equations.
The next step is based on an observation by J. Henn \cite{Henn:2013pwa}:
If a transformation can be found that brings the system of differential equations to an
$\eps$-factorised form
\bq
\label{eps_factorised}
 d I & = & \eps A\left(x\right) I,
\eq
where the only dependence on the dimensional regularisation parameter $\eps$ is through the explicit prefactor on the right-hand side,
a solution in terms of iterated integrals is straightforward.
This assumes that boundary values are known. These however constitute a simpler problem.
Often they can be obtained rather easily from regularity conditions.
This reduces the problem of computing a Feynman integral to finding an appropriate transformation 
to bring the differential equation into the form of eq.~(\ref{eps_factorised}).
It is an open question for which Feynman integrals such a transformation exists.
In this talk we show that a transformation does exist for the equal-mass banana integrals for all loop numbers $l$.
Our findings support the conjecture that a transformation to an $\eps$-factorised form can be found for all Feynman integrals.
Additional non-trivial support for this conjecture comes from refs.~\cite{Adams:2018yfj,Bogner:2019lfa,Muller:2022gec,Giroux:2022wav,Jiang:2023jmk,Dlapa:2022wdu,Gorges:2023zgv}.

Let us now look at the mathematical side: 
We have a vector bundle,
where the vector space in the fibre is spanned by the master integrals $I = (I_1, ..., I_{N_F})$.
The base space is parametrised by the coordinates $x=(x_1, ..., x_{N_B})$, which are the kinematic variables the Feynman integrals depend on.
More precisely, the master integrals $I_1(x), \dots, I_{N_F}(x)$ can be viewed
as local sections, and for each $x$ they define a basis of the vector space in the fibre.
In addition, the vector bundle is equipped with a flat connection
defined by the matrix $A$ made up of differential one-forms $\omega=(\omega_1, ..., \omega_{\NL})$.
On this vector bundle we have two operations at our disposal:
We may change the basis in the fibre $I'=U I$, leading to a new connection 
\bq
 A' & = & U A U^{-1} - U d U^{-1}.
\eq
Essentially, we look for a transformation $U$, such that the $\eps$-dependence factors out from the new connection $A'$.
In addition, we may perform a coordinate transformation $x_i'=f_i(x)$ on the base manifold.
If 
\bq
 A & = & 
 \sum\limits_{i=1}^{\NB} A_i dx_i
 \; = \;
 \sum\limits_{i=1}^{\NB} A_i' dx_i',
\eq
then $A_i'$ and $A_i$ are related by
\bq
 A_i' & = & \sum\limits_{j=1}^{\NB} A_j \frac{\partial x_j}{\partial x_i'}.
\eq


\section{Geometry}

Suppose we have a differential equation in an $\eps$-factorised form.
We can now ask if we can relate the base space to a space known from mathematics
by a suitable coordinate transformation.
Let's first look at an example.
Assume we have $(n-3)$ variables $z_1,\dots,z_{n-3}$ and differential one-forms
\bq
 \omega_k
 & \in &
 \left\{ d\ln\left(z_1\right), d\ln\left(z_2\right), \dots, d\ln\left(z_{n-3}\right),
 \right. \nonumber \\
 & &
 \left.
 \hspace*{1.5mm}
 d\ln\left(z_1-1\right), d\ln\left(z_2-1\right), \dots, d\ln\left(z_{n-3}-1\right),
 \right. \nonumber \\
 & &
 \left.
 \hspace*{1.5mm}
 d\ln\left(z_1-z_2\right), d\ln\left(z_1-z_3\right), \dots, d\ln\left(z_{n-2}-z_{n-3}\right)
 \right\}.
\eq
The iterated integrals on the base space with coordinates $z_1,\dots,z_{n-3}$
are multiple polylogarithms.
To see this, consider an integration path on the base space. 
The pull-back of the differential one-forms $\omega_k$ to the integration path leads to
differential one-forms of the type
\bq
 \omega^{\mathrm{mpl}} & = & \frac{d\lambda}{\lambda-c_j},
\eq
and iterated integrals of these differential one-forms are the 
multiple polylogarithms
\bq
G(c_1,...,c_k;\lambda) & = & \int\limits_0^\lambda \frac{d\lambda_1}{\lambda_1-c_1}
 \int\limits_0^{\lambda_1} \frac{d\lambda_2}{\lambda_2-c_2} ...
 \int\limits_0^{\lambda_{k-1}} \frac{d\lambda_k}{\lambda_k-c_k},
 \;\;\;\;\;\;
 c_k \neq 0.
\eq
To see the geometry consider first the configuration space of
$n$ distinct points on the Riemann sphere ${\mathbb C} \cup \{\infty\}$.
On a Riemann sphere we can perform M\"obius transformations 
and we mod out configurations that
are related by M\"obius transformations.
The space of equivalence classes of $n$ distinct points on the Riemann sphere modulo M\"obius transformations
is known as the moduli space ${\mathcal M}_{0,n}$ of a smooth complex algebraic curve of genus zero
with $n$ marked points.
The dimension of ${\mathcal M}_{0,n}$ is $(n-3)$, as we may use M\"obius transformations to fix three
points to prescribed positions, for example $z_{n-2}=0$, $z_{n-1}=1$ and $z_n=\infty$.
The requirement that the remaining points are distinct translates to
$z_i \notin \{0,1,\infty\}$ and $z_i \neq z_j$.
In the context of Feynman integrals the $z_i$ are usually 
functions of the kinematic variables $x$ and the arguments of the dlog-forms
are related to the Landau singularities.


\section{Elliptic curves}

\begin{figure}
\begin{center}
\includegraphics[scale=0.6]{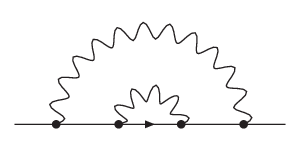}
\includegraphics[scale=0.6]{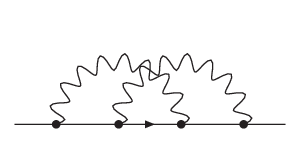}
\includegraphics[scale=0.6]{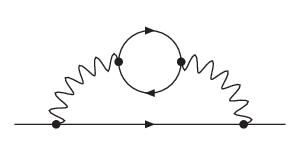}
\end{center}
\caption{
The Feynman graphs for the two-loop electron self-energy in QED.
}
\label{fig_self_energy}
\end{figure}

\begin{figure}
\begin{center}
\includegraphics[scale=0.6]{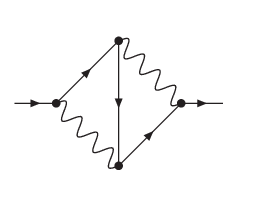}
\hspace*{6mm}
\includegraphics[scale=0.6]{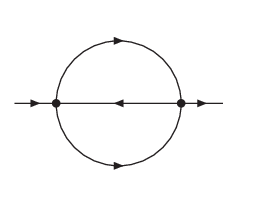}
\end{center}
\caption{
The kite graph (left) and the sunrise graph (right).
}
\label{fig_kite}
\end{figure}

It is well-known that not every Feynman integral can be expressed in terms of multiple polylogarithms.
Starting from two-loops, we encounter more complicated functions.
The next-to-simplest Feynman integrals involve an elliptic curve.
We do not have to go very far to encounter elliptic integrals in precision calculations:
The simplest example is the two-loop electron self-energy in QED \cite{Sabry:1962}:
There are three Feynman diagrams contributing to the self-energy,
as shown in fig~\ref{fig_self_energy}.
All master integrals are (sub-) topologies of the kite graph,
shown on the left in fig.~\ref{fig_kite}.
One sub-topology of the kite graph is the sunrise graph with three equal non-zero masses, shown on the right in 
fig.~\ref{fig_kite}.
The geometry of the sunrise graph is an elliptic curve.
This is most easily seen in the Feynman parameter representation.
The second graph polynomial defines an elliptic curve in Feynman parameter space:
\bq
 - p^2 a_1 a_2 a_3 + \left(a_1+a_2+a_3\right) \left(a_1 a_2 + a_2 a_3 + a_3 a_1 \right) m^2 & = & 0.
\eq
In analogy with the genus zero case we now consider the moduli space
${\mathcal M}_{1,n}$
of isomorphism classes of smooth complex algebraic curves of genus $1$ with $n$ marked points.
The dimension of ${\mathcal M}_{1,n}$ is $n$.
We have one coordinate which describes the shape of the elliptic curve. 
This coordinate is usually taken to be the modular parameter $\tau$, 
given as the ratio of two periods of the elliptic curve.
We may use translation to fix one marked point at a prescribed position, say $z_n=0$.
Thus, standard coordinates on ${\mathcal M}_{1,n}$ are $(\tau,z_1,...,z_{n-1})$.
Iterated integrals on ${\mathcal M}_{1,n}$ are iterated integrals of modular forms \cite{Adams:2017ejb}, 
elliptic multiple polylogarithms \cite{Broedel:2017kkb} and mixtures thereof.
These can be evaluated numerically within {\tt GiNaC} with arbitrary precision \cite{Walden:2020odh}.

\begin{figure}
\begin{center}
\includegraphics[scale=0.6]{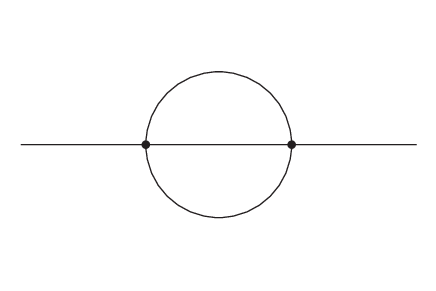}
\hspace*{3mm}
\includegraphics[scale=0.6]{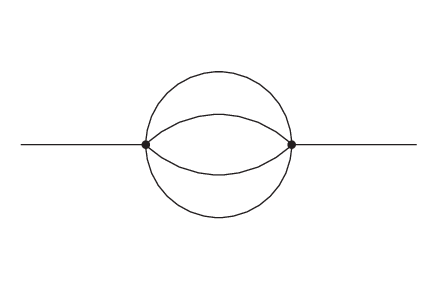}
\hspace*{3mm}
\includegraphics[scale=0.6]{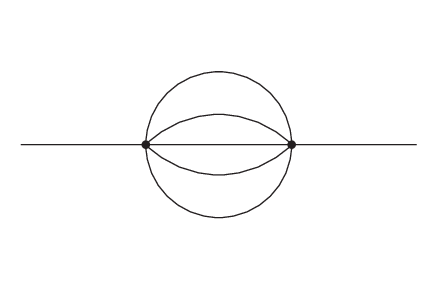}
\end{center}
\caption{
The banana graphs with two, three and four loops.
}
\label{fig_banana}
\end{figure}

With Feynman integrals related to algebraic curves of genus $0$ and $1$ well understood,
there is an obvious generalisation to algebraic curves of higher genus $g$, i.e.
iterated integrals on the moduli spaces ${\mathcal M}_{g,n}$.
Curves of genus two occur for example in non-planar double box integrals with internal masses \cite{Marzucca:2023gto}.
A second generalisation goes from 
curves to surfaces and higher dimensional objects.
This generalisation shows up in the banana graphs shown in fig.~\ref{fig_banana}, as we increase the number of loops $l$.
The geometry of the banana graphs with non-vanishing internal masses
are Calabi-Yau manifolds.


\section{Calabi-Yau manifolds}

A Calabi-Yau manifold of complex dimension $n$ 
is a compact K\"ahler manifold $M$ with vanishing first Chern class.
An equivalent condition is that $M$ has a K\"ahler metric with vanishing Ricci curvature.
This has been conjectured by Calabi \cite{Calabi:1954}
and was proven by Yau \cite{Yau:1978}.
Calabi-Yau manifolds come in pairs, related by mirror symmetry \cite{Candelas:1990rm}.
The mirror map relates a Calabi-Yau manifold $A$ to another Calabi-Yau manifold $B$ with Hodge numbers
$h^{p,q}_B = h^{n-p,q}_A$, as shown in fig.~\ref{fig_mirror}.
\begin{figure}
\begin{center}
\bq
\begin{array}{ccc}
 \begin{array}{ccccccc}
  & & & 1 & & & \\
  & & 0 & & 0 & & \\
  & 0 & & {\color{blue} h^{1,1}} & & 0 & \\
  1 & & {\color{orange} h^{2,1}} & & {\color{orange} h^{2,1}} & & 1 \\
  & 0 & & {\color{blue} h^{1,1}} & & 0 & \\
  & & 0 & & 0 & & \\
  & & & 1 & & & \\
 \end{array}
 & \hspace*{5mm} &
 \begin{array}{ccccccc}
  & & & 1 & & & \\
  & & 0 & & 0 & & \\
  & 0 & & {\color{orange} h^{2,1}} & & 0 & \\
  1 & & {\color{blue} h^{1,1}} & & {\color{blue} h^{1,1}} & & 1 \\
  & 0 & & {\color{orange} h^{2,1}} & & 0 & \\
  & & 0 & & 0 & & \\
  & & & 1 & & & \\
 \end{array}
 \\
 && \\
 \mbox{Calabi-Yau manifold $A$} && \mbox{Mirror manifold $B$} \\
\end{array}
 \nonumber 
\eq
\end{center}
\caption{
The mirror map relates the Calabi-Yau manifolds $A$ and $B$.
}
\label{fig_mirror}
\end{figure}
The $l$-loop banana integral with (equal) non-zero masses is related to a Calabi-Yau $(l-1)$-fold.
An elliptic curve is a Calabi-Yau $1$-fold, corresponding to the sunrise graph already discussed.

Our aim is to transform the system of differential equations for the equal-mass $l$-loop banana integral to an
$\eps$-factorised form. 
There are two key ingredients: 
The first ingredient is a change of variables from $y=-m^2/p^2$ (where $m$ denotes the internal mass and $p$ the external momentum)
to $\tau$, given by the mirror map.
More specifically, we consider the Picard-Fuchs operator on the maximal cut in two space-time dimensions.
The point $y=0$ is a point of maximal unipotent monodromy, and the solutions (called periods) 
can be ordered according to the Frobenius method in increasing powers of $\ln(y)$. 
The variable $\tau$ is given as the ratio of the single-logarithmic solution $\Frobeniusbasis_1$ 
by the holomorphic solution $\Frobeniusbasis_0$
of the Frobenius basis.

The second key ingredient is the 
special local normal form of a Calabi-Yau operator \cite{2013arXiv1304.5434B,2017arXiv170400164V}.
To motivate this form consider a sequence which starts as
\bq
\begin{array}{lc}
 l \; = \; 0: & 1, \\
 l \; = \; 1: & \theta, \\
 l \; = \; 2: & \theta \cdot \theta, \\
 l \; = \; 3: & \theta \cdot \theta \cdot \theta. \\
\end{array}
\eq
We would like to understand the general term at $l$ loops.
We first compute the $(l=4)$-term:
\bq
\begin{array}{lc}
 l \; = \; 4: & \theta \cdot \theta \cdot \frac{1}{Y_2} \cdot \theta \cdot \theta. \\
\end{array}
\eq
The general term at $l$ loops is given by
\bq
 \theta \cdot \frac{1}{Y_{l-1}} \cdot \theta \cdot \frac{1}{Y_{l-2}} \cdot \theta \cdot \frac{1}{Y_{l-3}} \cdot \ldots \cdot \frac{1}{Y_{3}} \cdot \theta \cdot \frac{1}{Y_{2}} \cdot \theta \cdot \frac{1}{Y_{1}} \cdot \theta,
\eq
and we have $Y_{1} = 1$
and the duality
$Y_{j} = Y_{l-j}$.
Up to seven loops explicit forms are
\bq
\begin{array}{lc}
 l \; = \; 5: & \theta \cdot \theta \cdot \frac{1}{Y_2} \cdot \theta \cdot \frac{1}{Y_2} \cdot \theta \cdot \theta, \\
 l \; = \; 6: & \theta \cdot \theta \cdot \frac{1}{Y_2} \cdot \theta \cdot \frac{1}{Y_3} \cdot \theta \cdot \frac{1}{Y_2} \cdot \theta \cdot \theta, \\
 l \; = \; 7: & \theta \cdot \theta \cdot \frac{1}{Y_2} \cdot \theta \cdot \frac{1}{Y_3} \cdot \theta \cdot \frac{1}{Y_3} \cdot \theta \cdot \frac{1}{Y_2} \cdot \theta \cdot \theta. \\
\end{array}
\eq
Here, $\theta$ is the Euler operator $\theta = \qbar \frac{d}{d\qbar}$ in the variable $\qbar=\exp(2\pi i\tau)$,
and the functions $Y_j$ are called $Y$-invariants.
The operators $N = \theta^2 \frac{1}{Y_{2}} \theta \frac{1}{Y_3} \dots \frac{1}{Y_3} \theta \frac{1}{Y_2} \theta^2$
are the special local normal form of Calabi-Yau operators,
and are related to Picard-Fuchs operators of Calabi-Yau Feynman integrals.
From the factorisation of $N$ in the variable $\tau$ (or $\qbar$) we may construct the $\eps$-factorised differential equation.
Note that non-trivial $Y$-invariants enter for the first time at $l=4$.


\section{The ansatz for the master integrals}

We now describe a method to derive an $\eps$-factorised differential equation for the $l$-loop equal-mass banana integrals.
This family has $(l+1)$ master integrals.
We set $D=2-2\eps$ and instead of $y=-m^2/p^2$ we work with the variable $\tau$ (or $\qbar$).
We construct master integrals
$M = \left(M_{0}, M_{1}, \dots, M_{l}\right)^T$,
which put the differential equation into an $\eps$-factorised form.
$M_0$ is proportional to the $l$-loop tadpole integral:
\bq
 M_{0}
 & = & 
 \eps^l I_{1 \dots 1 0}.
\eq
$M_{1}$ is constructed as follows:
$I_{1\dots 11}$ has a Picard-Fuchs operator $L^{(l)}$, the $\eps^0$-part $L^{(l,0)}$ is of the form
\bq
 L^{(l,0)} & = &
 \beta \theta^2 \frac{1}{Y_{l-2}} \theta \frac{1}{Y_{l-3}} \dots \frac{1}{Y_3} \theta \frac{1}{Y_2} \theta^2 
 \frac{1}{\Frobeniusbasis_0},
\eq
where $\Frobeniusbasis_0$ denotes the holomorphic solution of the Frobenius basis.
Note that $L^{(l,0)}$ is given by the special local normal form of a Calabi-Yau operator multiplied by a function $\beta$ from
the left and the function $1/\Frobeniusbasis_0$ from the right.
$L^{(l,0)}$ annihilates $I_{1\dots11}$ modulo $\eps$ and modulo tadpoles.
Furthermore, $M_1$ should start at order $\eps^l$.
This suggests
\bq
 M_{1}
 & = & 
 \frac{\eps^l}{\Frobeniusbasis_0} I_{1 \dots 1 1}.
\eq
The master integrals $M_2-M_l$ are constructed from an ansatz based on
Griffiths transversality
\bq
 M_{j}
 & = &
 \frac{1}{\Yinvariant_{j-1}} 
 \left[  
  \frac{1}{2 \pi i \eps} \frac{d}{d\tau} M_{j-1}
  - \sum\limits_{k=1}^{j-1} F_{(j-1)k} M_{k}
 \right]
 \;\;\;\;\;\; \mbox{for} \;\; j \ge 2,
\eq
with a priori unknown, but $\eps$-independent functions $F_{i j}(\tau)$.
The ansatz leads to the differential equation of the form
\bq
\label{coloured-diff_eq}
 \frac{1}{2 \pi i} \frac{d}{d\tau} M
 & = &
 \eps
 \left( \begin{array}{cccccccc}
 0 & 0 & 0 & 0 & 0 & \dots & 0 & 0 \\
 0 & \cellcolor{yellow} F_{11} & 1 & 0 & 0 & & 0 & 0 \\
 0 & \cellcolor{orange} F_{21} & \cellcolor{red} F_{22} & \Yinvariant_2 & 0 & & 0 & 0 \\
 0 & F_{31} & F_{32} & F_{33} & \Yinvariant_3 & & 0 & 0 \\
 \vdots & & & & & \ddots & & \vdots \\
 0 & \cellcolor{green} F_{(l-2) 1} & \cellcolor{lime} F_{(l-2) 2} & F_{(l-2) 3} & F_{(l-2) 4} & \dots & \Yinvariant_{l-2} & 0 \\
 0 & \cellcolor{cyan} F_{(l-1) 1} & F_{(l-1) 2} & \cellcolor{lime} F_{(l-1) 3} & F_{(l-1) 4} & \dots & \cellcolor{red} F_{(l-1) (l-1)} & 1 \\
 \ast & \ast & \cellcolor{cyan} \ast & \cellcolor{green} \ast & \ast & \dots & \cellcolor{orange} \ast & \cellcolor{yellow} \ast \\
 \end{array} \right) M,
\eq
where the first $l$ rows are in an $\eps$-factorised form.
We then determine the functions $F_{ij}$ such that the $(l+1)$-th row is also in $\eps$-factorised form.
The condition that in the $(l+1)$-th row only terms of order $\eps^1$ are present leads to 
differential equations and algebraic equations from self-duality.
Self-duality is the statement that entries of the same colour in eq.~(\ref{coloured-diff_eq}) are equal.
The equations for $F_{ij}$'s have a natural triangular structure
and can be solved systematically. 
This leads to a differential equation in $\eps$-factorised form:
\bq
 d M
 & = &
 \eps A M.
\eq


\section{Results and potential applications}

\begin{figure}
\begin{center}
\includegraphics[scale=0.35]{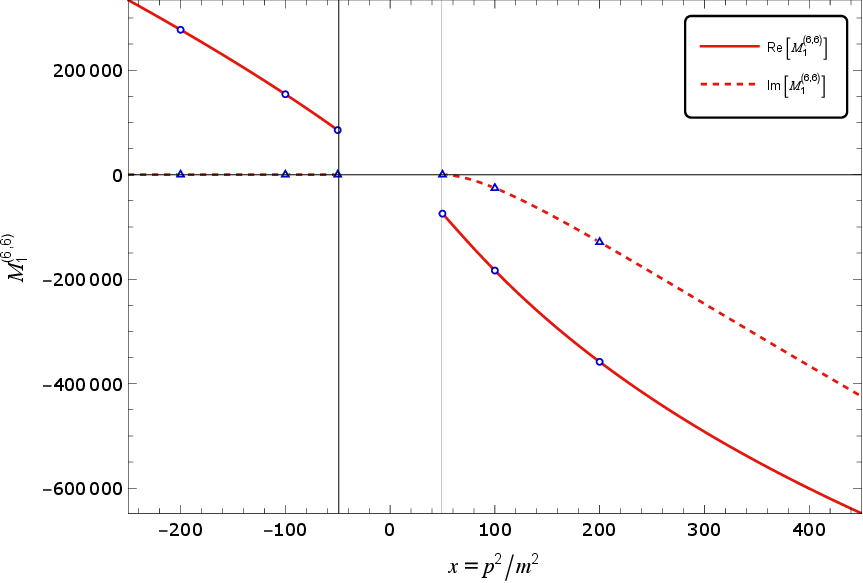}
\hspace*{5mm}
\includegraphics[scale=0.35]{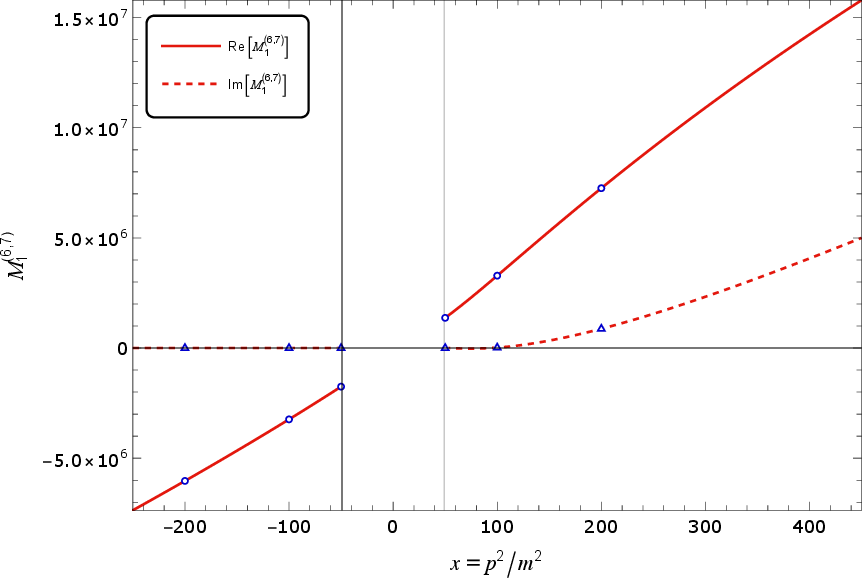}
\end{center}
\caption{
Results for the six-loop banana graph at order $\eps^6$ and $\eps^7$.
}
\label{fig_results}
\end{figure}

Fig.~\ref{fig_results} shows the real and imaginary part 
of the $\eps^6$-term and the $\eps^7$-term
of the six-loop equal-mass banana integral $M_1$ for $|p^2| > 49 m^2$ \cite{Pogel:2022vat}.
This is the region where the expansion around $y=0$ converges.
The results agree with results from {\tt pySecDec} \cite{Borowka:2017idc}.
The geometry of this Feynman integral is a Calabi-Yau $5$-fold.

We expect higher-dimensional Calabi-Yau Feynman integrals to be relevant for phenomenology.
For example, dijet production at N${}^3$LO will involve contributions related to a Calabi-Yau 2-fold,
top pair production at N${}^4$LO will involve contributions related to a Calabi-Yau 3-fold.
\begin{figure}
\begin{center}
\includegraphics[scale=0.6]{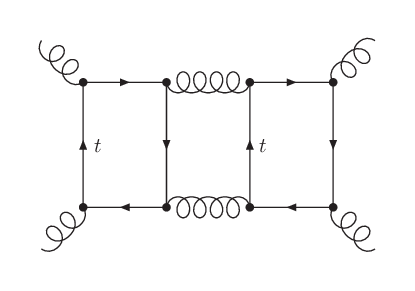}
\hspace*{4mm}
\includegraphics[scale=0.6]{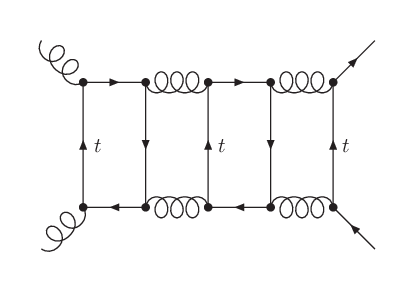}
\end{center}
\caption{
Feynman diagrams where we expect contributions related to higher-dimensional Calabi-Yau manifolds.
}
\label{fig_applications}
\end{figure}
The relevant Feynman diagrams involving internal top loops are shown in fig.~\ref{fig_applications}.

In summary, the results for the $l$-loop equal-mass banana integrals give
strong support for the conjecture that a transformation to an $\eps$-factorised differential equation 
exists for all Feynman integrals.


{\footnotesize
\bibliography{/home/stefanw/notes/biblio}
\bibliographystyle{/home/stefanw/latex-style/h-physrev5}
}

\end{document}